\documentclass[aps,prl,twocolumn,groupedaddress,showpacs]{revtex4}

\usepackage{graphicx}
\usepackage{dcolumn}
\usepackage{bm}

\begin{document}

\title{Quasi-Kondo Phenomenon due to Dynamical Jahn-Teller Effect}

\author{Takashi Hotta}

\affiliation{Advanced Science Research Center,
Japan Atomic Energy Research Institute,
Tokai, Ibaraki 319-1195, Japan}

\date{\today}

\begin{abstract}
A mechanism of non-magnetic Kondo effect is proposed
on the basis of a multiorbital Anderson model coupled with
dynamical Jahn-Teller (JT) phonons.
An electron system coupled dynamically with JT phonons
has a vibronic ground state with double degeneracy
due to clockwise and anti-clockwise rotational modes
with entropy of $\log 2$.
When a temperature is lower than a characteristic energy
to turn the rotational direction,
the rotational degree of freedom is eventually suppressed
and the corresponding entropy $\log 2$ is released,
leading to quasi-Kondo behavior.
We discuss a simple relation between the ``Kondo'' temperature
and the JT energy.
\end{abstract}

\pacs{75.20.Hr, 71.70.Ej, 75.40.Cx, 71.27.+a}


\maketitle


Forty years have passed since the pioneering work by Kondo
to solve the problem of resistance minimum phenomenon \cite{Kondo}.
It has been revealed that the heart of his solution is
the singlet formation from local magnetic moment
due to the coupling with conduction electrons \cite{Yoshida},
which is now called the Kondo effect \cite{Kondo40}.
The Kondo-like phenomenon is expected to occur more widely,
when a localized entity with internal degrees of freedom
is embedded in a conduction electron system.
Then, after the clarification of the Kondo effect
with magnetic origin,
lots of researchers have pursued a new mechanism of
Kondo phenomena with non-magnetic origin.

One possibility is to take into account explicitly
orbital degree of freedom of localized electron.
First, Coqblin and Schrieffer have derived exchange interactions
from the multiorbital Anderson model \cite{Coqblin}.
Then, the concept of multi-channel Kondo effect has been developed
on the basis of such exchange interactions \cite{Nozieres},
as a potential source of non-Fermi liquid phenomena
observed in some $f$-electron compounds.
Concerning the reality of the multi-channel Kondo effect,
Cox has pointed out the existence of the exchange interaction
in terms of quadrupole degree of freedom
in a cubic uranium compound
with non-Kramers doublet ground state \cite{Cox}.
There have been continuous experimental efforts
to detect such quadrupole Kondo effect.

Another possibility for the non-magnetic Kondo behavior
is to consider local phonons.
For instance, in an adiabatic approximation,
Einstein phonons provide a double-well potential
for an electron system, leading naturally to a two-level system,
in which Kondo has first considered a possibility of
non-magnetic Kondo behavior \cite{Kondo2}.
It has been shown that this two-level Kondo system exhibits
the same behavior as the magnetic Kondo effect \cite{Vladar}.
Further efforts to include the low-lying levels of local phonon
has been done intensively to unveil the quasi-Kondo behavior
in electron-phonon systems \cite{Miyake}.

Recently, heavy-fermion phenomena in Pr- and Sm-based
filled skutterudites have been experimentally found
\cite{Sato,Sanada}.
To understand such exotic properties of these compounds,
the off-center motion of rare-earth atom in the pnictogen cage
has been focused both from experimental \cite{Goto} and
theoretical viewpoints \cite{Miyake}.
In particular, the off-center motion has been suggested
to have degenerate $E_{\rm g}$ symmetry \cite{Goto}.
Inspired by this suggestion, we envision a scenario that
Jahn-Teller (JT) effect plays a crucial role for the problem
of Kondo-like behavior in filled skutterudites.
From a conceptual viewpoint, this is a challenge to understand
quasi-Kondo behavior by including both orbital and phonon
degrees of freedom.

In this Letter, we analyze a multiorbital Anderson model
with the local coupling term between $f$ electrons and JT phonons.
Since such a coupling is believed to exist
in Pr-based filled skutterudites,
we focus on a case for $n$=2,
where $n$ is the local $f$-electron number,
corresponding to Pr$^{3+}$ ion.
Using a numerical technique, we evaluate orbital susceptibility,
entropy, specific heat, and phonon displacement.
Then, we find that orbital fluctuations are enhanced in a vibronic state
with double degeneracy due to rotational modes of clockwise and
anti-clockwise directions, leading to residual entropy of $\log 2$.
When a temperature becomes lower than a characteristic energy
to turn the rotational direction,
the specific direction disappears due to the average
over a long enough time and the entropy $\log 2$ is released,
leading to Kondo-like behavior.
We also discuss a simple formula
for the ``Kondo'' temperature of this phenomenon.


Here we introduce the multiorbital Anderson model on the basis
of a $j$-$j$ coupling scheme for filled skutterudite compounds
\cite{Hotta1,Hotta2,Hotta3}.
The main conduction band is constructed from $p$ electrons of
pnictogen \cite{Harima1},
expressed by $a_{\rm u}$ with $\Gamma_7$ symmetry
in terms of the $j$-$j$ coupling scheme.
Note that $f$ electrons in $e_{\rm u}$ orbital with $\Gamma_8$
symmetry is localized. Then, the model is given by
\begin{eqnarray}
  H=\sum_{\bm{k}\sigma}
  \varepsilon_{\bm{k}} c_{\bm{k}\sigma}^{\dag} c_{\bm{k}\sigma}
  +\sum_{\bm{k}\sigma}
  (V c_{\bm{k}\sigma}^{\dag}f_{{\rm c}\sigma}+{\rm h.c.})
  +H_{\rm loc},
\end{eqnarray}
where $\varepsilon_{\bm{k}}$ is the dispersion of conduction electron,
$c_{\bm{k}\sigma}$ is the annihilation operator of conduction electron
with momentum $\bm{k}$ and spin $\sigma$,
and $f_{\gamma\sigma}$ is the annihilation operator of $f$ electron
on the impurity site with pseudospin $\sigma$ and orbital $\gamma$.
The orbital index $\gamma$ distinguishes
three kinds of the Kramers doublets,
two $\Gamma_8$ (``a'' and ``b'') and one $\Gamma_7$ (``c'').
$V$ is the hybridization between conduction and $f$ electrons
with $\Gamma_7$ symmetry.
Throughout this paper, we set $V$=0.05 and the energy unit is
taken as half of the conduction bandwidth,
which is about 1.4 eV \cite{Harima2}.
To adjust the local $f$-electron number as $n$=2,
we appropriately set the chemical potential of $f$ electron.

The local $f$-electron term $H_{\rm loc}$ consists of three parts
as $H_{\rm loc}$=$H_{\rm CEF}$+$H_{\rm ee}$+$H_{\rm eph}$,
where $H_{\rm CEF}$ denotes the crystalline electric field (CEF)
potential term, $H_{\rm ee}$ is the Coulomb interaction term among
$f$ electrons, and $H_{\rm eph}$ indicates the coupling term between
$f$ electrons and local JT phonons.
The CEF potential term is given by
$H_{\rm CEF}$=
$\sum_{\gamma\sigma} B_{\gamma}
f_{\gamma\sigma}^{\dag}f_{\gamma\sigma}$,
where $B_{\gamma}$ is the energy level depending on orbital $\gamma$.
Since the CEF term is already diagonalized,
it is convenient to introduce a level splitting $\Delta$
between $\Gamma_7$ and $\Gamma_8$ as
$\Delta$=$B_{\Gamma_8}$$-$$B_{\Gamma_7}$.
To reproduce the quasi-quartet situation with $\Gamma_1$ singlet ground
and $\Gamma_4$ triplet excited states for Pr-based filled
skutterudites \cite{CEF},
we prefer a small positive value for $\Delta$ as $\Delta$=$10^{-5}$.

The Coulomb interaction term $H_{\rm ee}$ is given by \cite{Hotta4}
\begin{eqnarray}
  H_{\rm ee} \!=\!
  (1/2) \sum_{\gamma_1 \sim \gamma_4}\sum_{\sigma_1, \sigma_2}
  I^{\sigma_1,\sigma_2}_{\gamma_1, \gamma_2, \gamma_3, \gamma_4}
  f_{\gamma_1\sigma_1}^{\dag}f_{\gamma_2\sigma_2}^{\dag}
  f_{\gamma_3\sigma_2}f_{\gamma_4\sigma_1},
\end{eqnarray}
where the Coulomb integral $I$ in the $j$-$j$ coupling scheme
is expressed by Racah parameters,
$E_0$, $E_1$, and $E_2$, which are set as
$E_0$=5, $E_1$=2, and $E_2$=0.5.

In the electron-phonon term $H_{\rm eph}$,
the effect of $E_{\rm g}$ rattling is considered to be
included as the relative vibration of surrounding atoms.
We remark that localized $\Gamma_8$ orbitals with
$e_{\rm u}$ symmetry have linear coupling with JT phonons with
$E_{\rm g}$ symmetry, since the symmetric representation of
$e_{\rm u}$$\times$$e_{\rm u}$ includes $E_{\rm g}$.
Thus, $H_{\rm eph}$ is given by
\begin{eqnarray}
  \label{Eq:Heph}
  H_{\rm eph} &=& g (Q_2 \tau_x + Q_3 \tau_z)
  +(P_2^2+P_3^2)/2 \nonumber \\
  &+&(\omega^2/2)(Q_2^2+Q_3^2) 
  + b (Q_3^3-2Q_2^2Q_3),
\end{eqnarray}
where $g$ is the electron-phonon coupling constant,
$Q_2$ and $Q_3$ are normal coordinates for $(x^2-y^2)$-
and $(3z^2-r^2)$-type JT phonons, respectively,
$P_2$ and $P_3$ are corresponding canonical momenta,
$\tau_{x}$=
$\sum_{\sigma}(f_{{\rm a}\sigma}^{\dag}f_{{\rm b}\sigma}
+f_{{\rm b}\sigma}^{\dag}f_{{\rm a}\sigma})$,
$\tau_{z}$=
$\sum_{\sigma}(f_{{\rm a}\sigma}^{\dag}f_{{\rm a}\sigma}
-f_{{\rm b}\sigma}^{\dag}f_{{\rm b}\sigma})$,
$\omega$ is the frequency of local JT phonons,
and $b$ indicates the cubic anharmonicity.
Here the reduced mass for JT modes is set as unity.
It is convenient to introduce non-dimensional
electron-phonon coupling constant $\alpha$ and
the anharmonic energy $\beta$ as $\alpha$=$g^2/(2\omega^3)$
and $\beta$=$b/(2\omega)^{3/2}$, respectively.


In this paper, to analyze the Anderson model, we exploit
a numerical renormalization group (NRG) method \cite{NRG},
in which momentum space is logarithmically discretized
to include efficiently the conduction electrons
near the Fermi energy.
In actual calculations, we introduce a cut-off $\Lambda$
for the logarithmic discretization of the conduction band.
Due to the limitation of computer resources,
we keep $m$ low-energy states.
Throughout this paper, we set $\Lambda$=5 and $m$=3000.
Note that the temperature $T$ is defined as
$T$=$\Lambda^{-(N-1)/2}$ in the NRG calculation,
where $N$ is the number of the renormalization step.
The phonon basis for each JT mode
is truncated at a finite number $N_{\rm ph}$,
which is set as $N_{\rm ph}$=25 in this paper.


Since the dynamical JT phonons are coupled with orbital
fluctuations, we discuss orbital susceptibility to
understand the dynamical JT effect on electronic properties.
For the purpose, we evaluate the susceptibility for
corresponding multipole operator $X_{\Gamma \gamma}$,
where $X$ denotes the multipole symbol and
$\Gamma$ is the irreducible representation
with $\gamma$ to distinguish degenerate representations,
due to the proper combination of three angular momentum operators
$J_a$ ($a$=$x$, $y$, and $z$) for $j$=5/2.
We note that orbital operators, $\tau_{z}$ and $\tau_{x}$,
correspond to 3g quadrupoles,
$O_{{\rm 3g}u}$=$(2J_z^2-J_x^2-J_y^2)/2$
and
$O_{{\rm 3g}v}$=$\sqrt{3}(J_x^2-J_y^2)/2$,
respectively, while another orbital operator,
$\tau_{y}$=$-i\sum_{\sigma}
(f_{{\rm a}\sigma}^{\dag}f_{{\rm b}\sigma}
-f_{{\rm b}\sigma}^{\dag}f_{{\rm a}\sigma})$,
is expressed by 2u octupole,
$T_{\rm 2u}$=$\sqrt{15} \, \overline{J_xJ_yJ_z}/6$,
where the bar denotes the operation to take all possible
permutations in terms of cartesian components \cite{Shiina}.
Then, we evaluate three kinds of orbital susceptibility
$\chi_{a}$ for $\tau_{a}$ operators
within the standard linear response theory.
We also evaluated entropy $S_{\rm imp}$
and specific heat $C_{\rm imp}$ for $f$ electrons
and average displacements,
$\langle Q_i \rangle$ and $\sqrt{\langle Q_i^2 \rangle}$,
for $i$=2 and 3.


Before proceeding to the discussion on the effect of JT phonons,
we briefly review the results for $n$=2 without JT phonons.
As for details, readers should refer Ref.~\cite{Hotta3}.
From the NRG results for multipole susceptibilities,
we have found significant quadrupole fluctuations
in addition to magnetic (dipole and octupole) ones
at low temperatures.
This result has been intuitively understood from the local
electron configuration in the balance between Coulomb interaction
and CEF potential \cite{Hotta1,Hotta2,Hotta3}.
For $\Delta$$>$0, the local $f$-electron ground state is
$\Gamma_1$ singlet composed of two $\Gamma_7$ electrons.
The first excited state is $\Gamma_4$ triplet formed by $\Gamma_7$
and $\Gamma_8$ electrons, while the second excited state is
$\Gamma_5$ including a couple of $\Gamma_8$ electrons.
Even if $\Gamma_1$ singlet is ground state,
there exist $\Gamma_4$ and $\Gamma_5$ triplet states
with small excitation energy in Pr-based filled skutterudites.
In fact, we observe residual entropy $\log 3$ at low temperatures.
Since these triplet states carry quadrupole moments,
there remain quadrupole fluctuations in addition to magnetic ones.


Now we include the effect of dynamical JT phonons, but
without cubic anharmonicity.
In Figs.~1(a) and 1(b), we show orbital susceptibilities, entropy,
and specific heat for $\omega$=0.3 and $\alpha$=0.5.
As mentioned above, due to the balance between Coulomb interaction
and CEF potential, triplet state exists near
the $\Gamma_1$ singlet ground state.
Here we note that the $\Gamma_4$ triplet state is JT active,
since one electron is in the degenerate $\Gamma_8$ orbitals.
Since the energy of the excited triplet state becomes lower
due to the JT effect,
we obtain a vibronic ground state with residual entropy $\log 2$,
in which $\Gamma_8$ electron is tightly coupled with JT phonons.
In the vibronic state for $n$=2, spin degree of freedom is
suppressed in the non-Kramers states,
while dynamical JT phonons enhance orbital fluctuations,
as intuitively understood from the form of $H_{\rm eph}$.
Then, three kinds of orbital fluctuations for $\tau_x$, $\tau_y$,
and $\tau_z$ significantly remain in the low-temperature region.

\begin{figure}[t]
\includegraphics[width=1.0\linewidth]{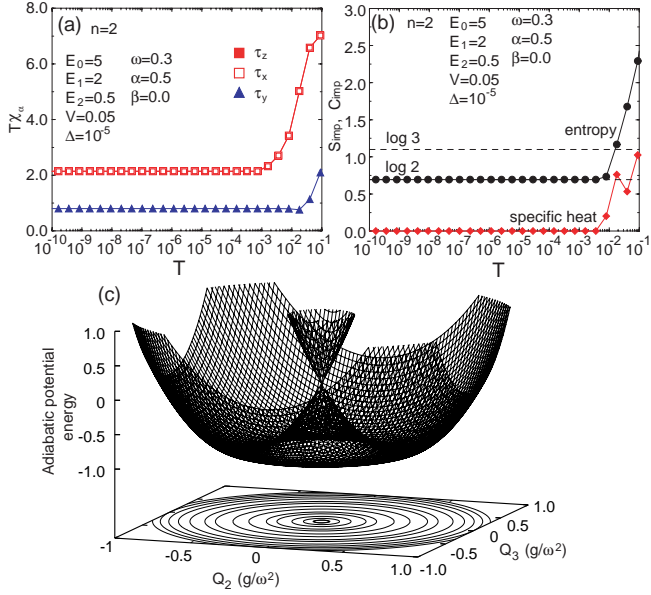}
\caption{(a) $T\chi_a$ and (b) $S_{\rm imp}$ and $C_{\rm imp}$
vs. temperature for $n$=2, $\Delta$=$10^{-5}$, $\omega$=0.3,
$\alpha$=0.5, and $\beta$=0.
(c) Local electron potential in an adiabatic approximation for
$\beta$=0.}
\end{figure}

Let us here discuss the nature of the vibronic state.
For the purpose, it is useful to see the electron potential
in an adiabatic approximation, as shown in Fig.~1(c),
although in actuality, the potential is not static,
but it dynamically changes to follow the electron motion.
For $\beta$=0, the potential is continuously degenerate
along the circle of the bottom of ``sombrero''.
Thus, there should exist double degeneracy for rotational modes
along clockwise and anti-clockwise directions,
as we observe the entropy of $\log 2$ in Fig.~1(b).
When a temperature becomes lower than a characteristic energy
$T^*$, which is related to a time scale to turn
the direction of rotational JT modes,
the residual entropy $\log 2$ should be eventually released,
leading to Kondo-like behavior,
since the specific rotational direction disappears
due to the average over the long enough time.
However, we cannot see such behavior
in the present temperature range,
because $T^*$ is considered to be very low.

\begin{figure}[t]
\includegraphics[width=1.0\linewidth]{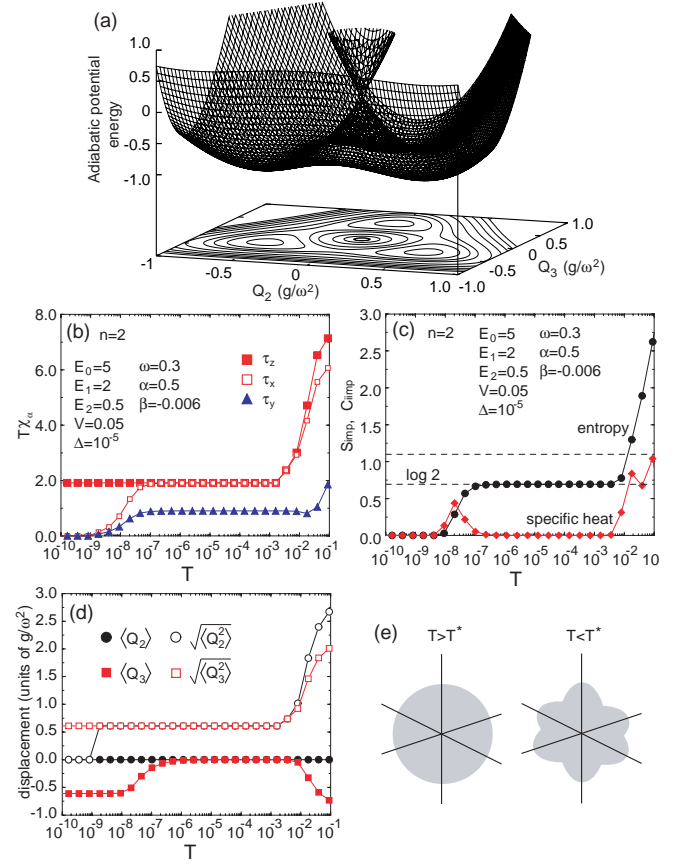}
\caption{(a) Local electron potential in an adiabatic approximation
for $\beta$$<$0.
(b) $T\chi_a$ ($a$=$x$, $y$ and $z$),
(c) $S_{\rm imp}$ and $C_{\rm imp}$,
and (d) $\langle Q_i \rangle$ and $\sqrt{\langle Q_i^2 \rangle}$
($i$=2 and 3) vs. temperature for $n$=2, $\Delta$=$10^{-5}$,
$\omega$=0.3, $\alpha$=0.5, and $\beta$=$-0.006$.
(e) Schematic views for displacements for $T$$>$$T^*$ and $T$$<$$T^*$.}
\end{figure}

In order to detect the quasi-Kondo behavior,
we increase $T^*$ by decreasing the excitation energy
concerning the rotational JT mode.
Such an energy is closely related to the static JT energy
$E_{\rm JT}$ in the adiabatic approximation.
In order to decrease effectively $E_{\rm JT}$ by keeping
the coupling strength, we include cubic anharmonicity.
As shown in Fig.~2(a), for $\beta$$<$0,
three potential minima appear in the bottom of the potential.
Since the rotational mode should be due to the quantum tunneling
among three potential minima at low temperatures,
the frequency is effectively reduced in the factor of
$e^{-\delta E/\omega}$,
where $\delta E$ is the potential barrier.
Then, we expect to observe the quasi-Kondo behavior
even in the present temperature range, when we include the
effect of cubic anharmonicity.


In Fig.~2(b), we show orbital susceptibilities for
$\omega$=0.3, $\alpha$=0.5, and $\beta$=$-0.006$.
In the present unit, $\delta E$ is about 100 K.
For $10^{-7}$$\alt$$T$$\alt$0.01, we observe orbital properties
similar to those in Fig.~1(a), since the vibronic ground state
also appears in this temperature range.
However, with decreasing temperature,
we observe the lift of degeneracy in orbital fluctuations and
the gradual decrease of $T\chi_x$ and $T\chi_y$ around at $T$=$10^{-8}$.
Then, only the $\tau_z$ fluctuation remains at low temperatures.
As shown in Fig.~2(c), at the crossover temperature $T^*$,
we actually observe the release of entropy $\log 2$ and
a clear peak in the specific heat,
suggesting the Kondo-like behavior,
even though $T^*$ is still low.

In order to understand what happens in this quasi-Kondo phenomenon,
let us show the temperature dependence of average displacements.
For $T^*$$\alt$$T$$\alt$0.01, we observe
$\sqrt{\langle Q_2^2 \rangle}$=$\sqrt{\langle Q_3^2 \rangle}$$\ne$0
and $\langle Q_2 \rangle$=$\langle Q_3 \rangle$=0.
Namely, there is no average displacement and the vibration is
always isotropic, as schematically shown in Fig.~2(d).
However, after the release of $\log 2$ entropy for $T$$<$$T^*$,
we obtain
$\sqrt{\langle Q_3^2 \rangle}$=$|\langle Q_3 \rangle|$$\ne$0
and $\sqrt{\langle Q_2^2 \rangle}$=$\langle Q_2 \rangle$=0,
indicating that the vibration becomes anisotropic
due to the selection of the $Q_3$-type mode,
as shown in Fig.~2(d).
Then, the degeneracy in orbital fluctuations is lifted and
only the $\tau_z$ fluctuation remains at low temperatures.

\begin{figure}[t]
\includegraphics[width=7cm]{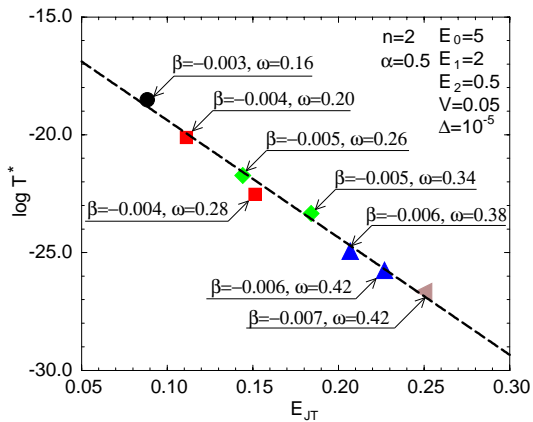}
\caption{
Characteristic temperature $T^*$ vs. $E_{\rm JT}$.
Dashed line is a guide for eyes.
Size of the symbol indicates the order of the precision
of $T^*$ in the calculation.}
\end{figure}


Finally, let us intuitively discuss a simple formula
for the characteristic temperature $T^*$.
For such a purpose, it is convenient to consider
the exchange interaction $J$ for this case.
Since $J$ should be related to the process of disappearance
of JT distortion, it is roughly expressed as
$J$$\propto$$V^2/E_{\rm JT}$,
where $E_{\rm JT}$ is given by
$E_{\rm JT}$=$\alpha \omega$+$8|\beta|\alpha^{3/2}$.
Here we recall the expression for the Kondo temperature,
$T_{\rm K}$=$W{\rm exp}(-1/J\rho_0)$,
where $W$ is the bandwidth of conduction electrons and
$\rho_0$ is the density of states at the Fermi level.
Then, we deduce the relation for $T^*$ as
$\log T^*$$\propto$$-E_{\rm JT}$.
In order to test the validity of this relation, in Fig.~3,
we plot $\log T^*$ vs. $E_{\rm JT}$ for several parameter sets,
where $T^*$ is defined as a temperature
showing a peak in $C_{\rm imp}$.
Within the errorbars, we can conclude that
the relation of $\log T^* \propto -E_{\rm JT}$ holds
for the quasi-Kondo phenomenon.
A similar expression for the Kondo temperature was obtained
in the context of odbital Kondo effect \cite{Nagaosa}.
Note, however, that this relation cannot be applied to all parameter
regions, since it is simply deduced in the adiabatic approximation for
JT phonons.
We also note that the effect of Coulomb interactions is ignored,
but it appears as an additional effect
for the renormalized JT energy \cite{Nagaosa,Hotta5} and thus,
the expression for $T^*$ does not change qualitatively.


In summary, we have numerically analyzed the multiorbital Anderson model
coupled with dynamical JT phonons.
We have observed the remarkable Kondo-like phenomenon induced by
dynamical JT effect and discussed a simple formula for ``Kondo''
temperature.
As for reality in actual materials \cite{Clougherty},
rattling motion in filled skutterudites provides
dynamical JT phonons coupled to degenerate orbitals.
Thus, it is interesting to pursue possible relevance of
the quasi-Kondo phenomenon to exotic heavy fermion behavior
in filled skutterudites.


The author thanks K. Kubo, H. Onishi, and M. Yoshizawa for discussions.
He is supported by the Japan Society for the Promotion of Science
and by the Ministry of Education, Culture, Sports, Science,
and Technology of Japan.
The computation in this work has been done using the facilities
of the Supercomputer Center of Institute for Solid State Physics,
University of Tokyo.


\end{document}